\documentclass[twocolumn]{aastex63}

\usepackage{xspace}

\newcommand{\rg}{$\,r_{\rm g}$}
\newcommand{\degree}{$^{\circ}$}

\newcommand\rxj{{QSO~J1131$-$1231 \xspace}}
\newcommand\ika{{Fe K$\alpha$ \xspace}}

\usepackage{amsmath}
\shorttitle{Modeling Inflows, Outflows, and Moving Coronal Plasma}
\shortauthors{H. Krawczynski}

\begin{document}
\input{journals.inp}
\title{Generally Applicable Formalism for Modeling the Observable Signatures of Inflows, Outflows, and Moving Coronal Plasma Close to Kerr Black Holes}

\author{Henric\,Krawczynski} \affil{Washington University in St. Louis, 
Physics Department, 
McDonnell Center for the Space Sciences, and 
the Center for Quantum Sensors,
1 Brookings Dr., CB 1105, St. Louis, MO 63130} 
\correspondingauthor{Henric Krawczynski, krawcz@wustl.edu}

\begin{abstract}
We present a generally applicable formalism for modeling the emission, absorption, reflection, and reprocessing of radiation 
by moving plasma streams close to a Kerr Black hole.
The formalism can be used to investigate the observational signatures of a wide range of phenomena, including: 
(i)\,the reflection of coronal X-ray radiation off plasma plunging from the inner edge 
of a black hole accretion disk towards the black hole;
(ii)\,the reflection of coronal X-ray emission off the upper layers of
a geometrically thick accretion flow;
(iii)\,the illumination of the accretion disk by a corona moving with
relativistic velocities towards or away from the accretion disk;
(iv)\,the emission from a jet forming close to the black hole.
After introducing the general relativistic treatment, we show the results for
a fast wind forming close to a Kerr black hole. The approach presented here can be 
used to model X-ray spectral, timing, reverberation, and polarization data.
\end{abstract}
\keywords{high energy astrophysics, X-ray astronomy, 
stellar mass black holes, supermassive black holes, inflows, outflows, winds}

\section{Introduction}
In this paper, we show how to model the emission, absorption, and reprocessing of radiation 
by inflows, outflows, or moving coronal plasma close to Kerr black holes.
The formalism is well suited to be used in general relativistic ray tracing codes 
\citep[e.g.][]{2005ApJS..157..335L,2009ApJ...701.1175S,2010ApJ...712..908S,
2012ApJ...754..133K,2013MNRAS.430.1694D,2015MNRAS.449..129W,
2017ApJ...850...14B,2017ApJ...851...57C,2018AA...619A.105T,2019ApJ...875..148Z}.
Such codes are likely to continue to play an important role for the analysis and
interpretation of the data from current and upcoming X-ray missions.
General Relativistic (Radiation) MagnetoHydroDynamic (GR(R)MHD) 
simulations have made enormous progress over the last decade 
\citep[see e.g.][and references therein]{2019ApJ...873...71K,2019ApJS..243...26P,2020MNRAS.tmp..707L}.
However, the predicted X-ray energy spectra will depend on a number 
of parameters  describing the background spacetime (i.e.\ the black hole mass and spin),
the properties of the accreted plasma (e.g.\ the accretion rate, the structure of the accreted 
magnetic field, and the elemental composition of the accreted plasma) that impact the 
accretion flow dynamics and thus the emitted energy spectra.
It seems unlikely that a sufficiently large number of GR(R)MHD simulations can be run 
(each individual simulation taking substantial time on a supercomputer) to enable  
the direct comparison of the simulated GR(R)MHD results with the X-ray data of many different 
astrophysical sources. More likely, the GR(R)MHD results need to be used to inform simpler treatments, 
such as the raytracing codes mentioned above. The latter can then be used in turn to simulate a 
sufficiently large number of configurations to allow for a quantitative comparison of the 
simulated and observed data. The work described in this paper can be used to add emitting, 
absorbing, reflecting or Comptonizing inflows, outflows, or relativistically moving coronal 
plasma to such raytracing codes.  

Several authors studied the impact of a moving corona on the illumination of the accretion disk and 
the shape of the reflected \ika line emission \citep{1997MNRAS.290L...1R,1999ApJ...510L.123B,2007ApJ...664...14F,2012MNRAS.424.1284W,2013MNRAS.430.1694D}.
These studies either considered flat spacetime geometries \citep{1997MNRAS.290L...1R,1999ApJ...510L.123B}, 
or were limited to the motion of the emitting gas along the spin axis of the black hole \citep{2007ApJ...664...14F,2012MNRAS.424.1284W,2013MNRAS.430.1694D}.
For notable exceptions see \citep{2002MNRAS.332L...1H,2004ApJS..153..205D}.
In contrast, the approach presented in this paper can be used to parameterize gas moving along any streamline or surface.
Application of the results of this paper include the phenomenological modeling of inflows and outflows.
A number of physics motivated wind and jet models have been put forward in the literature.
Most of the models however, are limited to describing the plasma motion far away 
from the black hole where the spacetime is approximately flat and a Newtonian approach 
is warranted \citep[e.g.][and references therein]{1976MNRAS.176..465B,1982MNRAS.199..883B,1999MNRAS.303L...1B,2010ApJ...723L.228F,2014ApJ...789...19H,
2014ARA&A..52..529Y,2015ApJ...814...87M,Waters_2016,2017MNRAS.465.2873N,2019ARA&A..57..467B,2020ApJ...899...80A}.
In contrast, the formalism discussed here can be used all the way to the event horizon.   

The rest of the paper is structured as follows. We describe the formalism in 
Section \ref{modeling}. As an example, we model the emission reflecting off an accretion disk and a fast
wind being launched close to a black hole in Section \ref{example}. We conclude with a discussion of 
other applications of the formalism in Section \ref{discussion}.
\section{General Relativistic Description of Plasma Moving Towards or Away from Kerr Black Holes}
\label{modeling}
\subsection{Flow Geometry and Kinematics}
\label{m1}
We use the Kerr metric in Boyer Lindquist coordinates $x^{\mu}\,=\,(t,r,\theta,\phi)$.
We employ geometric units, and set $M\,=\,1$. The spin parameter $a$ can take values 
between -1 (maximally counter-rotating) and 1 (maximally rotating). 
We assume that the flow can be parameterized by the functions $r(P)$ and $\theta(P)$ 
of $N$ parameters $P=\left\{p1, p2, ..., p_N\right\}$ with the parameter $p_1$ increasing
monotonically along the plasma streamlines. In the simplest case, the flow can be approximated to 
have the geometry of an axially symmetric surface, e.g., a dense wind, or the photosphere of a 
geometrically thick accretion flow. In that case, the parameter $p_1$ suffices to define the surface.

As an example, we will consider a conical outflow launched at $r=r_{\rm l}$ into the upper hemisphere 
with half opening angle $\alpha$.
In cylindrical coordinates ($\rho$,$z$), the outflow can be described by the equation:
\begin{equation}
z(\rho)\,=\,(\rho-r_{\rm l})/\tan{\alpha}
\end{equation}  
for $\rho\ge r_{\rm l}$. 
The corresponding Boyer Lindquist coordinates are:
\begin{eqnarray}
r(P)&=&\,\sqrt{\rho^2+z(\rho)^2}\\
\theta(P)&=&\arctan{(z(\rho)/\rho)}.
\end{eqnarray}
The two functions $r(P)$ and $\theta(P)$ with $P=\left\{p_1\right\}$, $p_1=\rho$
fully determine the geometry of the flow. One could model a 3-D flow by adding 
another parameter, e.g.\ $p_2\,=\,r_{\rm l}$, so that $p_2$ marks adjacent surfaces.

We will now derive an expression for the four velocity of the
flow {\bf v} given two additional functions: 
(i)~the velocity of the flow $\beta(P)$ relative to a local 
Zero Angular Momentum Observer (ZAMO) in units of the speed of light, 
and (ii)~the angular velocity of the flow $\Omega(P)$.
The parameterization of $\beta(P)$ and $\Omega(P)$ can be chosen freely, 
i.e. one can use the results from a GR(R)MHD simulation or other physically 
motivated prescriptions, or use a phenomenological parametrization with 
fitting parameters that need to be inferred from confronting 
the model predictions  with experimental data.

For our example, we assume that the flow co-rotates with the accretion plasma 
at ($r=r_{\rm l}$, $\theta=\pi/2$) and accelerates such that the Lorentz factor 
$\gamma\,=\,(1-\beta^2)^{-1/2}$ relative to the ZAMO is given by:
\begin{equation}
\label{e:gamma}
\gamma(P)=\gamma_{\rm l}+\frac{\left(\gamma _{\infty }-\gamma_{\rm l}\right) \arctan \left({z(\rho)}/{z_0}\right)}{{\pi }/{2}}.
\end{equation}
Here, $\gamma_{\rm l}$ is the Lorentz factor of matter in a Keplerian orbit at 
($r=r_{\rm l}$, $\theta=\pi/2$) as measured by a local ZAMO (see Equation (\ref{gl})),  
and $z_0\,=\,$20\rg\,gives the vertical length scale over which the flow accelerates.
In our example, we assume an asymptotic velocity of $\beta_{\infty}\,=\,0.75$ 
in units of the speed of light, corresponding to the Lorentz factor 
$\gamma_{\infty}=(1-\beta_{\infty}^{\,2})^{-1/2}\,\approx\, 1.51$.
We will determine $\Omega(P)$ from the assumption that the flow co-rotates with the accretion 
disk at the launching radius $r\,=\,r_{\rm l}$  and subsequently conserves its angular momentum.

We use the notation from the classical paper by \citet{1972ApJ...178..347B} to work out the kinematics. 
The ZAMO's four velocity {\bf u} has  contravariant components $u^{\mu}\,=\, (u^t,0,0,u^{\phi})$. 
The condition of vanishing angular momentum at infinity  
$L\,=\,u_{\phi}\,=g_{\phi \mu}u^{\mu}\,=0$ implies:
\begin{equation}
\omega\,\equiv\,u^{\phi}/u^{t}\,= 
-{g_{\phi t}}/{g_{\phi \phi }}.
\label{e:omega}
\end{equation}
The normalization condition 
\begin{equation}
u^2=-1 
\end{equation}
leads to
\begin{eqnarray}
u^t&=&\sqrt{A/\Sigma\Delta}
\label{e:zamo}
\end{eqnarray}
with
\begin{eqnarray}
A&\equiv&(r^2+a^2)^2-a^2\Delta\, {\rm sin}^2\theta,
\label{e:b1}
\\
\Delta&\equiv&r^2-2r+a^2,{\rm~and}
\label{e:b2}\\
\Sigma&\equiv&r^2+a^2 {\rm cos}^2\theta.
\label{e:b3}
\end{eqnarray}

We will show that the four velocity of the flow 
can be expanded into three vectors:    
\begin{eqnarray}
{\bf v}\,=\,a\,{\bf u}+b\,\partial_{\phi}+c\,{\bf t}
\label{e:exp}
\end{eqnarray}
with the expansion coefficients $a$, $b$, and $c$ to be determined. 
The superposition of the four velocity {\bf u} of the ZAMO and
the tangent vector $\partial_{\phi}$ along the $\phi$-direction 
allows us to describe orbital motion with angular velocities
above ($b>0$) or below ($b<0$) the angular velocity of the 
local ZAMOs. The vector {\bf t} describes the motion of the outflow along its $r$ and $\theta$ coordinates:
\begin{eqnarray}
{\bf t}\,=\,\dot{r}\,\partial_{r}+\dot{\theta}\,\partial_{\theta} \end{eqnarray}
with $\dot{r}\,=\,dr/dp_1$ and $\dot{\theta}\,=\,d\theta/dp_1$.
We will derive simple expressions for the expansion coefficients $a$, $b$ and $c$ as functions of 
$\beta(P)$ (or $\gamma(P)$) and $\Omega(P)$.

The angular velocity measured by a distant observer is $\Omega\,=\,v^{\phi}/v^t\,=\,\omega+\frac{b}{a \,u^t}$.
The parameter $b$ is thus given by:
\begin{eqnarray}
    b&=&a\,u^t(\Omega-\omega). \label{e:b}
\end{eqnarray}

The remaining parameters $a$ and $c$ follow from the
normalization condition $v^2=-1$ and the requirement
that the flow velocity measured by a ZAMO is $\beta$.
We perform the calculations with Bardeen's ZAMO tetrad:
\begin{eqnarray}
{\bf e}_{\tilde{t}}&\equiv&{\bf u},\\
{\bf e}_{\tilde{r}}&\equiv&\frac{1}{\sqrt{g_{rr}}}\,\partial_r,\\
{\bf e}_{\tilde{\theta}}&\equiv&\frac{1}{\sqrt{g_{\theta\theta}}}\,\partial_{\theta},\\
{\bf e}_{\tilde{\phi}}&\equiv&\sqrt{\frac{\Sigma}{A}}\frac{1}{\sin{\theta}}\,\partial_{\phi},\label{e:e3}
\end{eqnarray}
where the tildes mark quantities related to the ZAMO frame.
Denoting the contravariant components of {\bf v} in the ZAMO frame as ($v^{\tilde{t}},v^{\tilde{r}},v^{\tilde{\theta}},
v^{\tilde{\phi}}$), the normalization condition can be evaluated in the ZAMO frame:
\begin{equation}
v^2\,=\,-1\,=(v^{\tilde{t}})^{\,2}+(v^{\tilde{r}})^{\,2}+(v^{\tilde{\theta}})^{\,2}+(v^{\tilde{\phi}})^{\,2}.
\label{e:norm}
\end{equation}
The velocity of the flow along the 
$\tilde{j}^{\rm th}$ spatial basis vector 
measured by the ZAMO is: 
\begin{equation}
\beta^{\tilde{j}}\,=\,v^{\tilde{j}}/v^{\tilde{t}}
\end{equation}
for $\tilde{j}\,=\,\tilde{r},\,\tilde{\theta},\tilde{\phi}$. Combining the equation:
\begin{equation}
\beta^2\,=\,
(\beta^{\tilde{r}})^{\,2}+
(\beta^{\tilde{\theta}})^{\,2}+
(\beta^{\tilde{\phi}})^{\,2}
\end{equation}
with Equation~(\ref{e:norm}) gives:
\begin{equation}
(v^{\tilde{t}})^{\,2}\,=\,\frac{1}{1-\beta^2},
\end{equation}
which recovers the fact that $-v^{\tilde{t}}$ 
is the Lorentz factor $\gamma$ of the flow in the ZAMO frame.
We can get the ZAMO components of {\bf v} from dotting
{\bf v} with the tetrad vectors:
\begin{eqnarray}
v^{\tilde{t}}&=&\gamma\,=\,
-{\bf v}\cdot{\bf e}_{(\tilde{t})}\,=\,a \label{e:a}\\
v^{\tilde{r}}&=&
{\bf v}\cdot{\bf e}_{(\tilde{r})}\,=\,
\sqrt{g_{rr}}\, c\,\dot{r}\\
v^{\tilde{\theta}}&=&
{\bf v}\cdot{\bf e}_{(\tilde{\theta})}\,=\,
\sqrt{g_{\theta\theta}}\, c\,\dot{\theta}\\
v^{\tilde{\phi}}&=&
{\bf v}\cdot{\bf e}_{(\tilde{\phi})}\,=\,
b\,(\partial_{\phi}\cdot {\bf e}_{(\tilde{\phi})})
\,=\,b\,g_{\phi\phi}\,e_{(\tilde{\phi})}^{\,\,\,\,\,\,\phi},\label{e:phi}
\end{eqnarray}
with $e_{(\tilde{\phi})}^{\,\,\,\,\,\,\phi}\,=\,\sqrt{\Sigma/A}/\sin{\theta}$ from
Equation\,(\ref{e:e3}).

Equation~(\ref{e:norm}) leads to:
\begin{equation}
    c\,=\pm\sqrt{\,\frac{\gamma^2-(v^{\tilde{\phi}})^2-1}{g_{rr}\,\dot{r}^2+g_{\theta\theta}\,\dot{\theta}^2}}\label{e:c}
\end{equation}
with $c>0$ for an outflow and $c<0$ for an inflow.
Equations (\ref{e:b}), (\ref{e:a}), and (\ref{e:c})
give all three expansion coefficients, and thus determine {\bf v} once $\gamma$ and $\Omega$ 
are given. 

In our example, we assume the flow co-rotates with the accretion disk at its launching radius $r_{\rm l}$ and, 
once launched, conserves its angular momentum per unit mass.
In the equatorial plane of the accretion disk, the flow moves thus with the Keplerian four velocity $u_{\rm l}^{\mu}\,=\,u_{\rm l}^t\,(1,0,0,\Omega_{\rm l})$ 
with $\Omega_{\rm l}\,=\,(a+r_{\rm l}^{3/2})^{-1}$,
$u_{\rm l}^t\,=\,\sqrt{-1/(g_{tt}+2\, g_{t\phi}\,\Omega_{\rm l}+g_{\phi\phi}\,\Omega_{\rm l}^2)}$,
and carries the angular momentum $L_{\rm l}\,=\,(g_{\phi t}+g_{\phi\phi}\Omega_{\rm l})\,u_{\rm l}^t$ per unit mass. 
Angular momentum conservation, $g_{\phi\mu}\,v^{\mu}\,=\,L_{\rm l}$,
leads to: 
\begin{equation} 
b\,=\,\frac{L_{\rm l}}{g_{\phi\phi}}.\label{e:bb}
\end{equation}
As before $a=\gamma$, and $c$ follows from Equation~(\ref{e:c}) with
$v^{\tilde{\phi}}\,=\,$ $L_{\rm l}\,e_{(\tilde{\phi})}^{\,\,\,\,\,\,\phi}$.
In the equatorial plane, co-rotation with the accretion disk requires $c\,=\,0$,
and thus:
\begin{equation}
    \gamma\,=\,\gamma_{\rm l}\,\equiv\,\sqrt{1+(L_{\rm l}\,e_{(\tilde{\phi})}^{\,\,\,\,\,\,\phi})^2}.
    \label{gl}
\end{equation}
\subsection{Transformation from Global Boyer Lindquist Coordinates into the Rest Frame of the Flow and Vice Versa}
Given the four velocity of the flow {\bf v} 
in BL coordinates, we can define a 
tetrad that allows us to convert tangent vectors 
from BL coordinates to a comoving
Lorentzian reference frame and back.
We denote quantities relating to this tetrad with a hat. 
We get the tetrad of the flow from ${\bf e}_{(\hat{t})} \,=\, {\bf v}$
and from Schmidt-orthogonalizing the three spatial seed vectors $\partial_r,\,\partial_{\theta},\,\partial_{\phi}$
with the help of the dot product:  
\begin{eqnarray}
{\bf e}_{(\hat{r})}\! &=& \!{\rm Norm}[\partial_r \!+\! (\partial_r \!\cdot\! {\bf v}) \,{\bf v} ]   \\
{\bf e}_{(\hat{\theta})}\! &=&\! {\rm Norm}[\partial_{\theta} \!+\! (\partial_{\theta} \!\cdot\! {\bf v})\,{\bf v}\!-\! (\partial_{\theta} \!\cdot\! {\bf e}_{(\hat{r})})\,{\bf e}_{(\hat{r})} ]\\
{\bf e}_{(\hat{\phi})} \!&=& \!{\rm Norm}[\partial_{\phi} \! +\! (\partial_{\phi} \!\cdot\! {\bf v})\,{\bf v} 
\!-\!(\partial_{\phi}\! \cdot\! {\bf e}_{(\hat{r})})\,{\bf e}_{(\hat{r})}\nonumber \\ &&
\hspace*{2.95cm}\!- (\partial_{\phi} \!\cdot\! {\bf e}_{(\hat{\theta})})\,{\bf e}_{(\hat{\theta})}] 
\end{eqnarray}
where ``Norm[]'' stands for normalizing the argument four vector to 1.
The coordinates of an arbitrary four vector 
${\bf k}=k^{\mu}\partial_{\mu}=k^{\hat{\mu}}{\bf e}_{(\hat{\mu})}$ can be converted from BL 
coordinates (no hat) to tetrad coordinates (hat)
with the equations:
\begin{eqnarray}
k^{\hat{t}}&=&-{\bf k}\cdot {\bf e}_{(\hat{t})} {\rm \, and} \label{e1}\\
k^{\hat{j}}&=&{\bf k}\cdot {\bf e}_{(\hat{j})} {\rm \,for\,}\hat{j}=\hat{r},\hat{\theta},\hat{\phi}.\label{e2}
\end{eqnarray}
The inverse transformation follows from the contravariant 
BL coordinates 
$e_{(\hat{\mu})}^{\,\,\,\,\,\,\mu}$ of the tetrad vectors
$e_{(\hat{\mu})}$:
\begin{eqnarray}
k^{\mu}&=&k^{\hat{\mu}}\,(e_{(\hat{\mu})})^{\mu},\label{e3}
\end{eqnarray}
where summation over $\hat{\mu}=\hat{t},\hat{r},\hat{\theta},\hat{\phi}$ is implied.
\subsection{Reflection off a 2-D surface}
In the example of reflection off an optically thick conical outflow, the reflector can be modeled as a surface with two spatial dimensions.
Integrating the geodesic, the ray tracing code checks in each integration step if the geodesic crossed the reflecting surface. 
If it does, the step size is adapted, so that the step ends exactly on the reflector.
If ($\rho_i, z_i$) and ($\rho_{i+1}, z_{i+1}$) denote the start and end points of the integration step in cylindrical coordinates,
and $z(\rho_i)$ and $z(\rho_{i+1})$ give the $z$-coordinates of the outflow at these two locations, respectively,
the step size correction factor $\eta$ is given in linear approximation by:
\begin{equation}   
\eta\,=\,\frac{z_i-z(\rho_i)}{(z_i-z(\rho_i))+(z(\rho_{i+1})-z_{i+1}))}.
\end{equation}
Equations (\ref{e2}) and (\ref{e3}) are then used to transform the photon wave vector and polarization vector 
into the comoving frame and back to simulate the scattering. 
\subsection{Reprocessing of radiation in a spatially extended flow}
In the more general case, photons can penetrate the plasma of the flow, and the latter needs
to be modeled as an object with three spatial dimensions, adding one more parameters 
to the parameterization of the flow (see Section \ref{m1}).
 Assuming the outflowing plasma is launched at proper time $\tau=0$ from the location $x^{\mu}(0)$
with a rest frame mass density $\rho(0)$ and moves with the four velocity {\bf v} from Equation (\ref{e:exp}),
we can infer the plasma position $x^{\mu}$ as function of proper time from:
\begin{equation}
x^{\mu}(\tau)\,=\,x^{\mu}(0)+
\int_{0}^{\tau}\,v^{\mu}(x^{\mu}(\tau'))\,d\tau'
\label{i1}
\end{equation}
The rest mass conservation law in differential form reads \citep[e.g.][]{1974ApJ...191..499P,2017grav.book.....M}:
\begin{equation}
\boldsymbol{\nabla}\cdot\left(\rho_0 {\bf v}\right)\,=\,0.
\end{equation}
Writing the four divergence in component form:
\begin{eqnarray}
\left(\rho_0\,v^{\mu}\right)_{\,\,;\mu}
&=& \rho_{0,\mu}\,v^{\mu}+\rho_{0}\,v^{\mu}_{\,\,;\mu}\\ 
&=& \frac{d\rho}{d\tau}+\rho_0\,v^{\mu}_{\,\,;\mu}=0
\end{eqnarray}
where ``,'' and ``;'' denote the partial and covariant derivatives, respectively. 
The plasma density along the streamline is then given by:
\begin{equation}
\rho_0(\tau)\,=\rho_0(0)\,-\int_{0}^{\tau}\,\rho_0(\tau')\,v^{\mu}_{\,\,;\mu}(\tau')\,d\tau'.
\label{i2}
\end{equation}
Combining Equations (\ref{i1}) and (\ref{i2}), the rest frame plasma density can be calculated 
everywhere in the flow based on a parametrization of its density at the launching sites.
Note that the treatment does not enforce energy and momentum conservation, as hydromagnetic forces might  
add or remove energy and momentum.  

The simulation of radiative processes can proceed as follows.  The ray tracing code checks for each integration step if 
a part or the entire integration step falls inside the flow. The step size should be adapted when entering or exiting the flow,
so that a new step starts at the boundary.
For an integration step starting at $x_i^{\,\,\mu}$ and ending at $x_{i+1}^{\,\,\,\,\,\,\mu}$, Equation (\ref{e2}) can be used to obtain 
the transformation from BL coordinates to plasma coordinates at the midpoint location $x_{\rm m}^{\,\,\mu}\,=(x_{i}^{\,\,\mu}+x_{i+1}^{\,\,\,\,\,\,\mu})/2$.
After transforming the displacement vector $\Delta x^{\mu}\,=x_{i+1}^{\,\,\,\,\,\,\mu}-x_{i}^{\,\,\mu}$ from BL coordinates
into $\Delta x^{\hat{\mu}}$ in the plasma frame, the proper spatial distance $\Delta l$ 
traversed in the integration step can be calculated from the spatial components of $\Delta x^{\hat{\mu}}$.
Using $\rho(x_{\rm m}^{\,\,\mu})$ and $\Delta l$ and the transformed wave and polarization vectors of the incoming photon beam in the 
plasma frame, the probability for a radiative interaction and the properties of the outgoing photon beam can be calculated. 
Following the calculation of $\Delta l$, the step size may need to be reduced and the step may need to be retaken 
to allow for proper integration of the relevant cross sections in sufficiently small steps.   
The treatment of tracking photons through a plasma described here, mirrors the treatment of 
tracking photons through coronas described in several general relativistic Comptonization codes  
\citep[e.g.][]{2010ApJ...712..908S,2017ApJ...850...14B,2019ApJ...875..148Z}. 
The treatment is of course not limited to simulate the Comptonization of radiation, 
but can be adapted to simulate absorption, reflection, or any other radiative processes.   
\subsection{Polarized radiation transport}
The treatment above details how to use a flow geometry, velocity and initial density description to derive the 
the four velocity, rest frame density, and transformation matrices everywhere within the flow.
The framework can be used to implement unpolarized and polarized radiation transport.
In the general relativistic ray tracing code {\tt xTrack} \citep{2012ApJ...754..133K,2016PhRvD..93d4020H,2017ApJ...850...14B,2019ApJ...870..125K,2020ApJ...889..111A},
the polarized radiation transport is based on tracking photon beams through the global coordinate system 
keeping track of a statistical weight $w$, the position $x^{\mu}$,  wave vector {\bf k}, 
the linear polarization fraction $p_{\rm l}$ and the relative circular polarization intensity $V/I$
\citep[two Lorentz invariants,][]{1972NPhS..240..161C},
and the linear polarization four vector {\bf f} (with $f^2=1$ and {\bf k}$\cdot${\bf f}=0) which 
gives the preferred direction of the electric field vector of the beam \citep{2017grav.book.....M}.
For rays propagating through vacuum, the code integrates the geodesic equation and 
parallel transports {\bf k} and {\bf f}. 

The effect of Faraday rotation, and the conversion of linearly to circularly polarized radiation 
and vice versa can be implemented in a two-step process \citep[see, e.g.,][]{2018MNRAS.475...43M}.
For each integration step, the code parallel transports {\bf k} and {\bf f}. 
Subsequently, the effect of the plasma and the quantum vacuum on the polarization of the beam is accounted for. 
This is accomplished by updating the polarization vector {\bf f} and polarization parameters $p_{\rm l}$ and $V/I$ as follows. 
The BL components of {\bf k} and {\bf f} at the end of the integration step are transformed into the plasma frame with Equation (\ref{e2}). 
Subsequently, the polarization parameters are packaged into a Stokes vector $I_j$:
\begin{eqnarray}
I_j=
\begin{pmatrix}
i_j\\
q_j\\
u_j\\
v_j
\end{pmatrix}
=
\begin{pmatrix}
1\\
p_{\rm l} \cos{2\psi}\\
p_{\rm l} \sin{2\psi}\\
V/I
\end{pmatrix}
\end{eqnarray}
with $\psi$ being the angle between a reference direction and the spatial part of {\bf f} in the plasma frame.   
The Stokes vector is then evolved according to: 
 \begin{equation}
I_{j+1}\,=\,I_j+\Delta l\,
\begin{pmatrix}
1& 0 & 0 & 0 \\
0& 0 & \rho_V & - \rho_U \\
0& -\rho_V & 0 & \rho_Q \\
0& \rho_U & -\rho_Q & 0 \\
\end{pmatrix}\,
I_j
\label{stokes}
\end{equation}
with $\Delta l$ being the proper distance of the integration step in the plasma frame (see previous section), 
and the $\rho_Q$, $\rho_U$ and $\rho_V$ are the rotation and conversion coefficients at the midpoint of the integration step.
The updated Stokes vector is then used to calculate {\bf f}, $p_{\rm l}$, and $V/I$ in the plasma frame, and in the BL frame. 
Rapid oscillations of the different Stokes vector components can be simulated by replacing Equation (\ref{stokes}) with 
analytically integrated expressions \citep{2018MNRAS.475...43M}.
For more general treatments of polarized radiation transport that allow for emission and absorption processes
as well as cross frequency transport see 
\citet{2004MNRAS.349..994B, 2012ApJ...752..123G,2016MNRAS.462..115D,2018MNRAS.475...43M,2020A&A...641A.126B,2020ApJ...897..148G,2020ApJ...888...94D} and references therein. 
\section{Example: simulation of the emission reflected by a fast outflow} 
\label{example}
As an example of the application of the formalism described above, we use the {\tt xTrack} code 
to predict the energy spectra resulting from the reflection of coronal lamppost emission 
\citep{1991A&A...247...25M} off a fast wind being launched close to a near-maximally spinning 
Kerr black hole with spin $a=0.98$.
We are interested in the model because the relativistic deboosting of the reflection off the 
far (receding) side of the outflow may explain the extremely redshifted emission lines 
found in individual {\it Chandra} observations of the gravitationally lensed quasar QSO~J1131$-$1231
\citep{2017ApJ...837...26C}. 
Driving such a wind would require a non-negligible fraction of the Eddington luminosity $L_{\rm Edd}$ (see Appendix).

We assume a lamppost corona located close to the spin axis of the black hole at the radial 
BL coordinate $r\,=\,25\,r_{\rm g}$, isotropically emitting a power law $dN/dE\propto E^{-\Gamma}$ 
with photon index $\Gamma=1.7$ in its rest frame. 
A geometrically thin, optically thick accretion disk extends 
from the Innermost Stable Circular Orbit (ISCO) at $r_1\,=r_{\rm ISCO}\,=\,$1.61\rg\, 
to the radial coordinate  $r_2\,=\,100\,r_{\rm g}$. 
At radial coordinate $r_{\rm l}\,=\,20\,r_{\rm g}$, the accretion flow launches 
the conical outflow with a half opening angle of $\alpha\,=\,$ 60\degree.

As mentioned above, the reflection of the coronal emission off the fast outflow is implemented by 
first transforming the photon packet's wave and polarization vectors from the BL coordinates 
into the rest frame of the outflowing plasma making use of Equations (\ref{e1}),\,(\ref{e2}).
Photons impinging on the disk or on the wind are reflected. 
We use the inclination dependent reflection spectrum from the Feautrier code XILLVER 
of \citet{2014ApJ...782...76G} with $\Gamma=1.7$ and ionization parameter log$(\xi)\,=$\,1.3 
to describe the reflected emission in the plasma frame.  The impinging photon packet's 0-component 
$k^0$ in the plasma frame is used to calculate the statistical weight of the reflected energy spectrum  
(weight $\propto$ $(k^0)^{\Gamma-1}$). 
The packet's wave vector is subsequently back-transformed into the global BL frame
with the help of Equation (\ref{e3}). Reflected photons impinging another time 
onto the disk or outflow are neglected.

\begin{figure}[thb]
 \begin{center}
    \includegraphics[width=5.4cm]{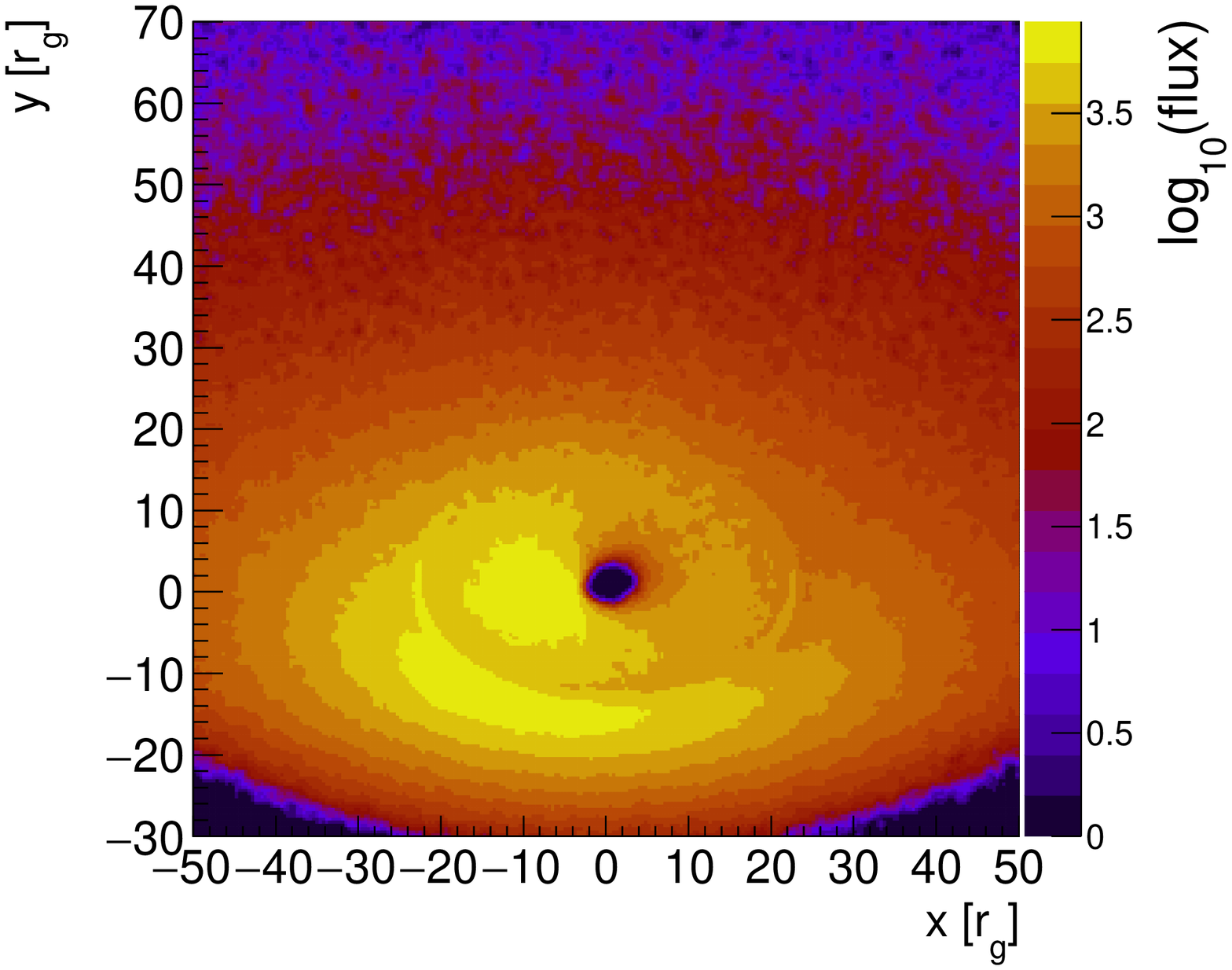}
    \includegraphics[width=5.4cm]{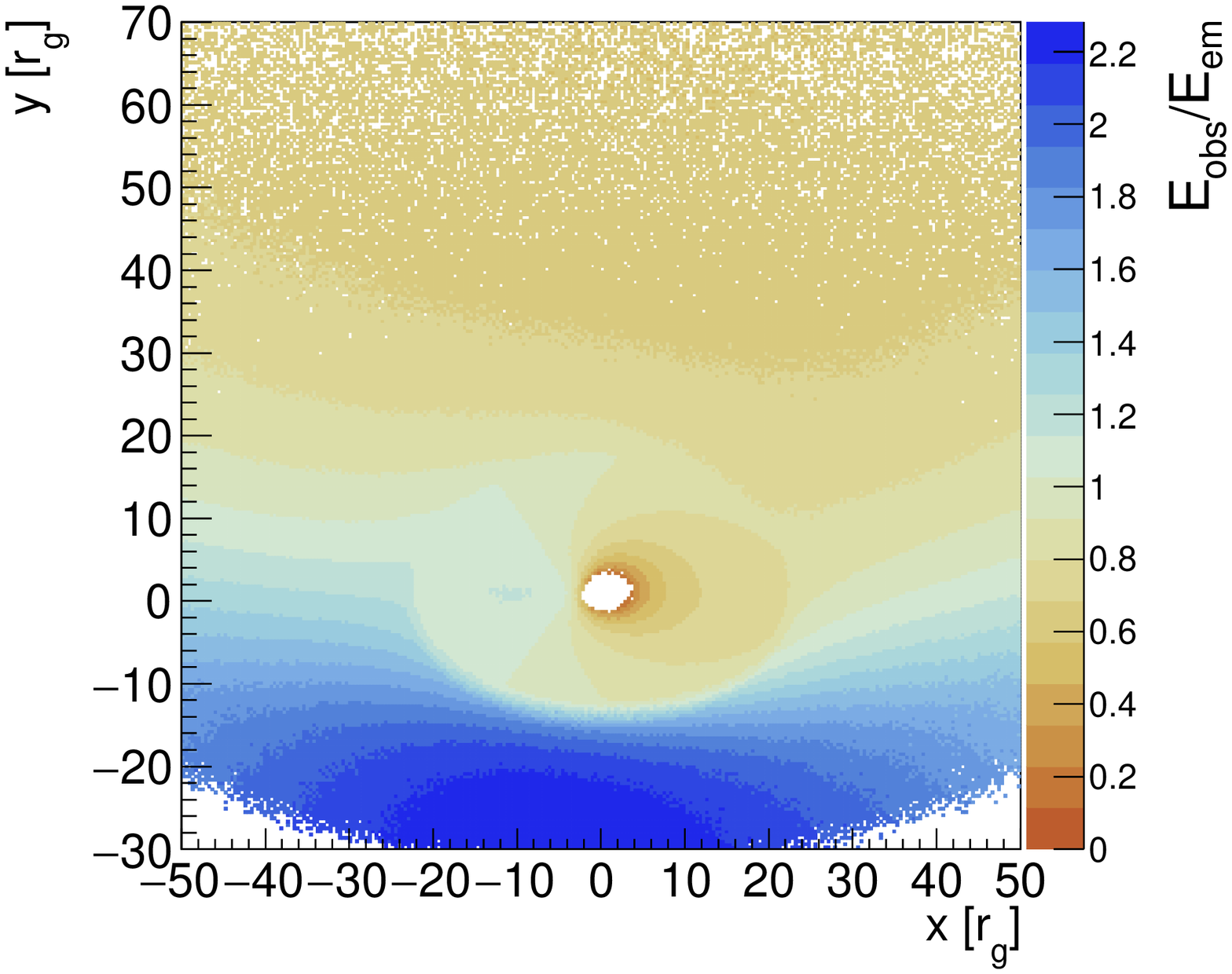}
    \includegraphics[width=4.7cm]{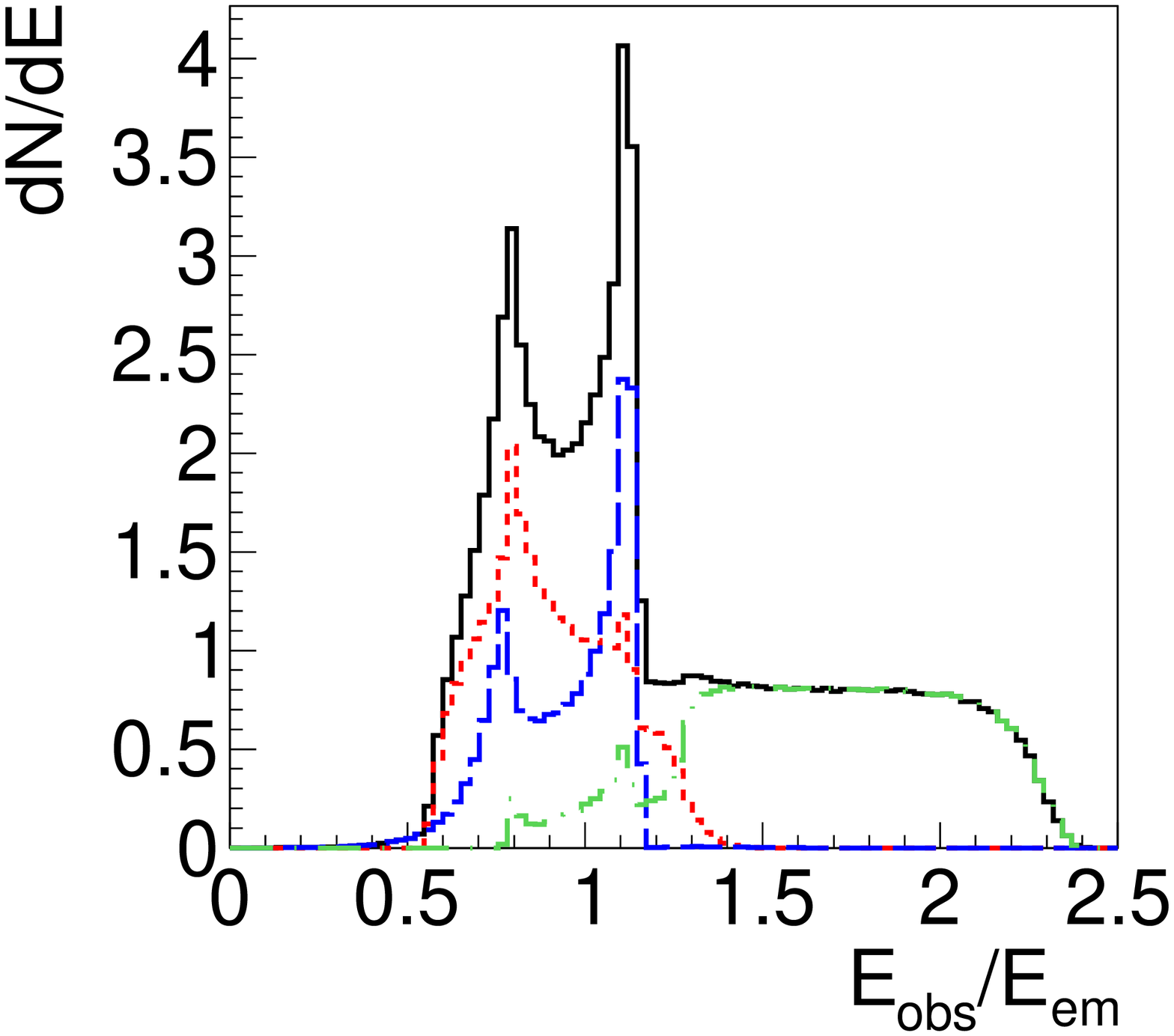}
    \end{center}

    \caption{Intensity (left, logarithmic scale) and relative change of the photon energy between emission
    in the plasma frame and observation (center and right panels) of the reflected emission of a mass accreting Kerr black hole ($a=0.98$) 
    driving a fast wind with an asymptotic velocity of 0.75\,$c$.
    The results are shown for an observer at an inclination of 50$^{\circ}$ relative to the spin axis of the black hole.
    In the right panel, the different line styles show 
    the entire reflected emission (black solid line), 
    and the emission reflected off 
    (A) the far side of the wind (red dotted line),
    (B) the accretion disk at radial coordinates $r\,<\,20\,r_{\rm g}$ (blue dashed line), and
    (C) the near side of the wind (green dash-dotted line).     \label{f:o1}}
\end{figure}

\begin{figure}[thb]
 \begin{center}
    \includegraphics[width=8cm]{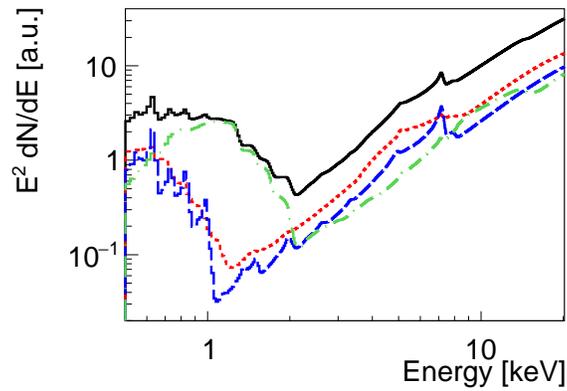}
    \caption{Spectral Energy Distribution (SED) of the reflected emission from the configuration of Figure \ref{f:o1}
    obtained with the {\tt xTrack} ray tracing code and the {\tt XILLVER} reflection code \citep{2014ApJ...782...76G}. 
    See Figure \ref{f:o1} for the description of the different line styles. 
    \label{f:o2}}
\end{center}
 \end{figure}

Figures~\ref{f:o1} and ~\ref{f:o2} shows the results from the simulation.
The configuration leads to three major emission components (see the different line styles in 
Figure~\ref{f:o1}, right panel, and Figure~\ref{f:o2}). 
Component (A) comes from the far side of the  conical outflow moving at an approximate angle 
$\theta_{\rm far}\,\approx\, i+\alpha=110$\degree relative to the observer. 
As the motion is mostly transverse to the line of sight,  
the emission is relativistically de-boosted by Doppler factors down to 
$\delta_{\rm far}=[\gamma_{\infty}(1-\beta_{\infty}\cos{\theta_{\rm far}})]^{-1}\approx\,$0.5, 
giving rise to emission between 3.5\,keV and 5\,keV. 
Component~(B) comes mostly from the inner $<20$\rg\,region of the 
accretion flow. The emission from Component~(C) comes from the near side
of the flow moving towards the observer at an approximate 
observer frame angle of $\theta_{\rm near}\,\approx\,\alpha-i=10$\degree.
The relativistic motion boosts the photon energy 
by Doppler factors up to 
$\delta_{\rm near}=[\gamma_{\infty}(1-\beta_{\infty}\cos{\theta_{\rm near}})]^{-1}\approx\,$2.5, giving rise to emission up to 16 keV.
Figure~\ref{f:o2} shows the resulting Spectral Energy Distribution (SED).
The SED exhibits a rather broad excess with a subdued peak at 4 keV coming from the far side of the wind, 
a pronounced peak at 7 keV coming mostly from the approaching side of the accretion disk, 
and rather smooth excess emission between 7 keV and 16 keV coming from the near side of the wind.
Fitting the model to actual data sets is  outside of the scope of this paper.
\section{Discussion}
\label{discussion}
This paper presents a formalism that can be used to model inflows, outflows, and moving coronal plasma 
close to Kerr black holes.  The formalism requires the flow velocity $\beta$ and angular velocity $\Omega$ 
of the plasma relative to a ZAMO. We show the results for an example where a generic parametrization 
is used for  $\beta$, and $\Omega$ is inferred from angular momentum conservation. 
For future applications one could envision to use GR(R)MHD simulations or other physical inflow or outflow models to 
parameterize $\beta$ and $\Omega$.

We have shown in Section \ref{example} that the approach can be used to model the 
reflection of coronal X-ray emission off an accretion disk and off a fast outflow. 
The self consistent modeling of the reflection off the two flow components makes it 
possible to predict not only the spectral shapes of the different reflection components, 
but also their relative intensities modulo uncertainties in the physical properties of the plasma.
The calculations can furthermore predict the absolute and relative polarization 
and time offsets of the reflection components. Modeling of suitable data sets can be 
used to constrain not only the black hole spin and inclination, but also the physical 
properties and geometries of the reflectors.

The formalism can be used to analyze a wide range of phenomena:
\begin{enumerate}
\item Although plasma reaching the Innermost Stable Circular Orbit will plunge into the black hole,
the emission cannot be neglected entirely 
\citep[see][]{2004ApJS..153..205D,2011ApJ...743..115N,2011MNRAS.414.1183K,2012MNRAS.424.2504Z}.
Recently, {\it NICER} and {\it NuSTAR} observations of the Galactic black hole X-ray binary 
MAXI J1820+070 show evidence for soft excess emission \citep{2020MNRAS.493.5389F}.
The authors speculate that magnetic stresses at the ISCO may explain the delayed infall and may power the emission.  
With suitable input from GR(R)MHD simulations, our formalism can be used to parametrize the 
motion of the plunging matter and to model the intensity and energy spectrum of the emission.
\item  The accretion disks of black holes accreting at super-Eddington rates are expected to be 
geometrically thick, and the reflection of coronal emission off the funnel walls should lead to observable signatures 
\citep[e.g.][]{2019BAAS...51c.352B,2019ApJ...884L..21T,2020MNRAS.496.2922M}.
Along these lines, \citet{2016Natur.535..388K} invoke the \ika emission from a relativistically moving upper accretion disk layer 
to explain the detection of a blue shifted iron lag from the tidal disruption event Swift J1644$+$57. 
The formalism introduced here can be used to quantitatively test models of the 
geometry and velocity of the flow by comparing the simulated with the predicted spectral and reverberation properties.
\item Observations of several Active Galactic Nuclei show strong evidence for a changing corona location or a changing corona geometry  
\citep[e.g.][]{2009MNRAS.394..427B,stu1246,2015MNRAS.449..129W,2015MNRAS.454.4440W}.
Our formalism can be used to ascertain not only the impact of different corona geometries 
but also the impact of moderately or highly relativistic motion of the coronal plasma \citep[see][]{1997MNRAS.290L...1R,1999ApJ...510L.123B,2007ApJ...664...14F}.
The emission from plasma moving relativistically towards the accretion disk can explain extremely 
high reflection ratios owing to the coronal emission being boosted in the direction of the disk.
Conversely, plasma moving relativistically away from the accretion disk can explain extremely 
low reflection ratios owing to the deboosting of the coronal emission.
The parameterization shown in this paper is not limited to describing motion along 
the spin axis of the black hole as in the earlier studies of \citet{2012MNRAS.424.1284W,2013MNRAS.430.1694D}.
\item The approach discussed here could also be used to simulate the emission, reflection, or reprocessing of radiation
by non-axis-symmetric geometries. \citet{2001PASJ...53..189K,2002MNRAS.332L...1H,2004ApJ...613..700F} 
for example consider the emission from accretion flows with spiral structure. 
Simulating non-axis-symmetric geometries will require to account for the azimuthal 
dependence of the flow geometry or the flow properties (e.g. the plasma density or the ionization degree).
\end{enumerate}
%
\section*{Appendix}
In this appendix, we give a rough estimate of the power required to drive an accretion disk wind as
described in Section \ref{example}. The physical scale of system depends on the black hole mass and we do the estimate here for
the $\sim 1.3\,\times\,10^8$ solar mass black hole of the quasar \rxj \citep{2006ApJ...649..616P,2006A&A...449..539S,2010ApJ...709..278D,2012A&A...544A..62S}. 
The projected neutral hydrogen column density of the winds needs to be $\sim 10^{23}$ ~cm$^{-2}$.
Assuming the photons impinge on the wind at the oblique angle $\alpha\,=\,60^{\circ}$, 
the Thomson optical depth is: 
\begin{equation}
\tau_{\rm es} \,=\, 1.2\,\frac{N_{\rm H}}{\cos{\alpha}}\,\sigma_{\rm T}\approx 16\%
\end{equation} 
with the factor 1.2 accounting for He atoms, and $\sigma_{\rm T}$ being the Thomson cross section.
The corresponding K-shell photo-ionization optical depth for photons of energy $E_{\gamma}$ is \citep{2003ApJ...592.1089B}:
\begin{equation}
\tau_{\rm Fe K}(E_{\gamma})\,=\,\frac{N_{\rm H}}{\cos{\alpha}}\,A_{\rm Fe}\,\sigma_K (E_{\gamma}).
\end{equation}
with $A_{\rm Fe}$ being the relative iron abundance, and $\sigma_K (E_{\gamma})$ 
the K-shell photoionization cross section from \citep{1995A&AS..109..125V}. 
For a solar system iron abundance of $2.82 \times 10^{-5}$, we get $\tau_{\rm Fe K}\,=$ 
19\% and 8\% for $E_{\gamma}=7.1$\,keV and $E_{\gamma}=10$\,keV, respectively.
Neglecting the impact of electron scattering before and after the photoionization,   
the probability of a photon of energy $E_{\gamma}>7.1$~keV prompting the emission 
of an \ika  photon into the hemisphere of the observer is given by \citep{2003ApJ...592.1089B}:
\begin{equation}
P_{\rm Fe K\alpha}(E_{\gamma})\,=\,
\frac{1}{2}\,\omega\,b\,(1-e^{-\tau_{\rm Fe K}(E_{\gamma})}).
\end{equation}
The factor 1/2 gives the probability of emitting the \ika photon into the hemisphere of the observer,
$\omega\,=\,0.342$ is the K fluorescence yield of iron, and $b\,=\,0.822$ is the K$\alpha$ to 
K$\beta$ branching ratio. 
The expressions gives a \ika yield of 2.5\% and 1.1\% for primary photons of 7.1\,keV and 10\,keV, respectively. 
If such a wind propagates at the distance  $r_{\rm W}\,=30\,r_{\rm g}$ from the black hole's 
spin axis with velocity $\beta$, the corresponding mass loss rate is in Newtonian approximation:
\begin{equation}
\dot{M}\,=\, \,N_{\rm H}\,\beta\,c\,2\pi\,30\,r_{\rm W}\,\frac{m_{\rm P}}{X_{\rm Sun}}
\end{equation}
with $X_{\rm Sun}\,\approx\,0.74$ being the solar hydrogen mass fraction, and $m_{\rm P}$ the proton mass.
Driving this outflow requires a mechanical luminosity of 
\begin{equation}
L_{\rm wind}\,=\, (\gamma-1) \dot{M} c^2
\end{equation}
with $\gamma=(1-\beta^2)^{-1/2}$. 
For $\beta=0.25$, 0.5, and 0.75, the mechanical luminosity is 3\%, 14\%, and 46\% of the 
Eddington Luminosity $L_{\rm Edd}$ of a 1.3$\times 10^8$ solar mass black hole.
Note that $L_{\rm wind}$ and $L_{\rm Edd}$ both increase linearly with black hole mass, so that 
$L_{\rm wind}/L_{\rm Edd}$ is independent of the black hole mass.
\section*{Acknowledgements}
The author would like to thank Q. Abarr, A. West, L. Lisalda, A. Hossen, F. Kislat, and M.~Nowak for joint discussions. 
Q.\,Abarr contributed a Cash-Karp integrator with adaptive step size to the ray tracing code, and 
F.\ Kislat accelerated the code substantially by removing redundant trigonometric calculations. 
HK thanks two referees for their excellent comments. HK acknowledges NASA for the support through the grants NNX16AC42G and 80NSSC18K0264.

\bibliographystyle{aasjournal}
\bibliography{outflow}
\end{document}